\documentclass[aps,nofootinbib,prx,twocolumn,groupedaddress]{revtex4-2}
\usepackage{graphicx}
\usepackage{subfig}
\usepackage{caption}
\usepackage{amsmath}
\usepackage{amssymb}
\usepackage{braket}
\usepackage{xcolor}
\usepackage{hyperref}
\usepackage{siunitx}
\hypersetup{
	colorlinks,
	linkcolor={red!50!black},
	citecolor={blue!80!black},
	urlcolor={blue!50!black}
}

\begin{document}
\title{Designer quantum reflection from a micropore}
\author{Romuald Kilianski}
\email[email: ]{r.kilianski.1@research.gla.ac.uk}
\author{Robert Bennett}
\affiliation{University of Glasgow,
University Avenue,
G12 8QQ, Glasgow}

\date{\today}

\begin{abstract}
We expand the theoretical toolbox for controllable quantum reflection by departing from a simple planar reflector. We introduce a circular hole (a micropore) of variable size, for which the electrostatic image potential can be exactly calculated. We combine this with two-dimensional simulations of wavepacket propagation at arbitrary angle of incidence to show that the quantum reflection probability can be tuned over a wide range of values.
\end{abstract}

\maketitle

\section{Introduction} 
Quantum reflection is a counter-intuitive effect in which an attractive surface-atom potential exhibits a repulsive behaviour towards an incoming matter-wave, i.e., the reflection occurs despite the absence of a classical turning point \cite{barrier1}. In other words, an atom that is accelerated towards a surface has a non-zero chance of reflection before coming into contact with it. This quintessentially wave behaviour is familiar from classical theories, i.e., describing wave propagation in inhomogeneous media \cite{brekhovskikh2012waves}, but it is its quantum realisation that has been the focus of research in recent decades.

 Quantum reflection has been a widely studied phenomenon since the inception of quantum mechanics and was first placed into the realm of atom-surface interactions by Jones and Devonshire \cite{lennard1936interaction}. In this context, the effect translates to an interference of probability waves; the unexpected outcome is that the reflection occurs at a threshold distance away from the surface. This effect was neatly demonstrated in a recent work \cite{Dimer}, where a helium dimer was non-destructively reflected before it could reach the surface at which the potential would have been strong enough to dissociate its weak bond.

A wealth of literature exists on the theoretical treatment of quantum reflection \cite{theo1,theo2,theo3,theo4,CP2}, some recent publications were concerned with reflection of atoms from rough surfaces \cite{CP1,HaraldJacoby}, some with antimatter (antihydrogen) reflecting off nanoporous materials \cite{gab2} and liquid helium \cite{antiHydroCrepin}, while others probed ultracold molecular collisions \cite{molecular}, and the effect of reflection in a cold Rydberg atomic gas with the use of single photons \cite{Photons}, and solitons \cite{Soliton}.  Since the seminal paper by Shimizu \cite{Shimizu2001}, experimental realisations of quantum reflection from a solid surface have become abundant \cite{exp1,exp2}. Some of the recent works include using quantum reflection to trap atoms in optical potentials \cite{trap}, the reflection of Bose-Einstein condensates \cite{bose,BEC_Wang_2021}, metrological applications via observation of diffraction orders \cite{difford} and tests of quantum vacuum \cite{qvacuum}. The diverse range of phenomena that can be probed via quantum reflection, makes it an exciting and versatile tool in atomic physics.   

The electromagnetic forces playing the key role in many realisations of quantum reflection belong to the class of phenomena collectively known as dispersion forces  \cite{Buhmann:2012cda}. They arise as a result of the field fluctuations between two objects that do not possess a permanent electric or magnetic dipole moment. Amongst them are interatomic van der Waals forces (vdW), initially proposed by London \cite{London}, and interactions between larger bodies, introduced by Casimir \cite{Casimir1948} and Lifshitz \cite{Lifshitz}. A third, mixed case describes the force between an atom or a molecule and a macroscopic object. It was first developed in the electrostatic regime by Lennard-Jones \cite{Lennard-Jones_s}, and then extended to the retarded distances by Casimir and Polder (CP) \cite{CP_original}. Different naming conventions for the specific dispersion forces exist in the literature, but all are on a fundamental level expressions of the fluctuations of the vacuum, as described by quantum electrodynamics (QED). We will follow the convention adopted in \cite{Buhmann:2012cda}, and refer to atom-body interactions --- that are of central importance in this work --- as CP forces\footnote{Some authors refer to any far-field dispersion interaction involving at least one atom as Casimir-Polder, while others use the convention employed here where Casimir-Polder refers to the interaction (at any distance) between an atom and macroscopic body.}. Moreover, different distance regimes impose limits on applicability of different theoretical descriptions of dispersion forces. Due to the finite speed of light, the information exchanged on scales much larger than the transition wavelength of an atom will suffer a phase delay, thus experiencing retardation effects. In this paper, we consider the short, non-retarded regime --- this is the domain of applicability of the electrostatic potential we will use.

Control of dispersion forces therefore allows control of quantum reflection. Research so far in this direction has mainly been centred around investigating the versatility of graphene as a material, enabling the control of the atom-surface potential. The works in \cite{graphene,graphene2, graphene3} explore the use of a magnetic field to alter the properties of a sheet of graphene, effectively changing the CP interaction potential. Some other investigations in the graphene-based systems include carrier doping \cite{grapheneDoping}, the application of mechanical strain \cite{grapheneStrain}, and plane stacking arrangements \cite{graphenePlanes}. Moreover, various experimental applications have already been realised \cite{grapheneExp,grapheneExp2,grapheneExp3}, paving the way for future uses in nanotechnology. Other materials in the graphene family have also been probed in the context of quantum reflection and its connection with topological phase transitions, stimulated by electric fields \cite{grapheneFamily}. The application of electric fields has equally been successful in controlling quantum reflection in silicon gratings--- outside of the realm of graphene--- as experimentally shown in \cite{grating}. 

In this work we choose to investigate the paradigmatic case of a perfectly reflecting plate, but with a twist. We introduce a circular hole (a micropore) to our metallic plate. By changing the diameter of the hole, we are able to reduce the potential gradient in its immediate neighbourhood, and in turn, gain control of the strength of reflection of, for example, a matter-wave passing through. Despite the presence of the micropore, which one might expect to allow full transmission of a highly-collimated matter-wave, the atom continues to reverse its motion at the threshold, albeit at a reduced rate relative to no hole. This apparent ``anti-tunnelling" event, where in the region of absence of a material surface one would expect the atom to propagate through, renders a curious addition to an already counter-intuitive quantum phenomenon. After incorporating the variable hole diameter, we extend the space of control parameters by including the angle of incidence and test their impact on the reflectivity. Due to the non-separability of the potential, we numerically solve the time dependent Schrödinger equation (TDSE), for $^{3}\mathrm{He}$ and Na, both modelled as a Gaussian pulse propagating towards a plate with a hole. 

This paper is organised as follows, we first discuss the conditions for the quantum reflection, considering a single degree of freedom. Secondly, in section \ref{procedure} we present the exact, two-dimensional, non-separable potential for a perfectly reflecting plate with a hole from \cite{EberleinZietal}. We then modify its domain to enable numerical simulations of quantum reflection by solving the TDSE using a spectral, split-step method. We then proceed to present the results in Section \ref{results}, showcasing the dependence of the reflectivity upon the hole diameter and angle of incidence. In the appendices we validate our algorithm for the case of normal incidence and examine the influence of the grid size on convergence.

\begin{figure}[h]
	\centering
	\includegraphics[width=1\linewidth]{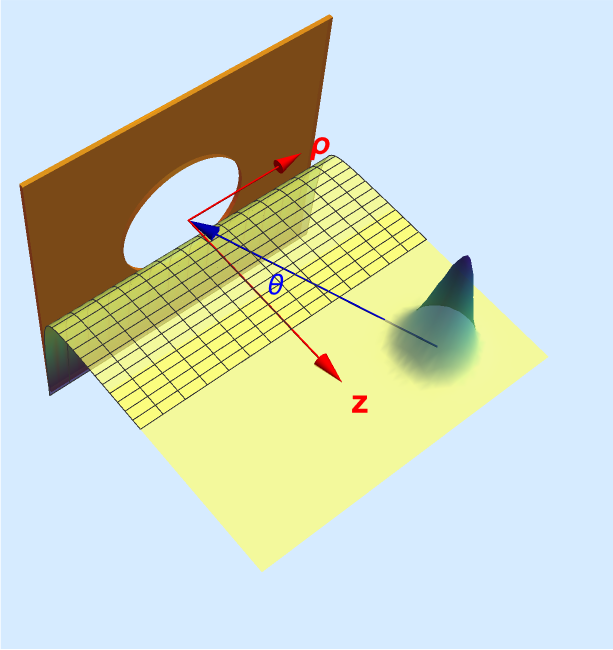}
	\caption{A schematic illustration of the setup. A pulse travels towards the plate at an incidence angle $\theta$, being influenced by an attractive potential $U(\rho,z;d)$.  The potential was originally derived in the cylindrical coordinates but we discard the $\phi$-dependent component due to the uniaxial invariance and treat $\rho$ and $z$  as Cartesian coordinates. }
	\label{fig:pwhgraphic}
\end{figure}
\section{Procedure}\label{procedure}
\subsection{Quantum reflection in 1D and 2D}
Quantum reflection has been studied overwhelmingly as a one-dimensional problem. The atom-surface forces between matter and a regular macroscopic object depend on the normal distance between them, resulting in consideration of only a single degree of freedom. The conditions for the quantum reflection in 1D are determined by the properties of the travelling matter-wave. If $U(x)$ is an arbitrary potential varying in the $x$ direction, $m$ is the particle's mass, and $k_{0}$ is its $k$-vector at $x\rightarrow \infty$ ($U\rightarrow 0$), the local wave vector of the particle $k$,
\begin{align}
k = \sqrt{k_{0}^{2}-2m U(x)/\hbar^{2}},
\end{align}
is required to change abruptly on the scale of its de Broglie wavelength $\lambda_{dB}$ for quantum reflection to occur \cite{Shimizu2001}. This significant change can be mediated by an interaction potential $U$ that grows rapidly as the atom approaches the surface. CP forces with their $1/r^{3}$ or $1/r^{4}$ dependence are therefore ideal for inducing quantum reflection. The majority of theoretical investigations into quantum reflection elude fully analytical treatment and rely on semiclassical approaches such as WKB approximation \cite{WKB1,WKB2}, however these are not applicable in higher dimensions for non-separable potentials \cite{Galiffi_2017_2D_sim}. Only until recently \cite{Galiffi_2017_2D_sim}, all efforts have been confined to the one dimensional case, in which a time-independent Schrödinger equation (TISE) is solved for a given potential, and the reflectivity is obtained as a ratio of amplitudes of counter-propagating waves. Quite understandably, a reliable method of solving a two-dimensional quantum reflection problem is a recent occurrence due to the computationally expensive nature of such setup. Inspired by the aforementioned work by Galiffi et al.~\cite{Galiffi_2017_2D_sim}, we apply a time-dependent approach to solve a pulse propagation problem in the vicinity of a perfectly reflecting plate with a hole. 
\subsection{Potential function}
The geometry we have chosen is a smooth metal plate with a hole in its centre, as shown in Fig.~\ref{fig:pwhgraphic}.
The exact electrostatic potential for this situation was calculated by Eberlein and Zietal \cite{EberleinZietal} by the means of a Kelvin transform \cite{KelvinTrans}. Defined in cylindrical coordinates, $U(\rho,\phi, z) $, and for a hole diameter $d$ their result can be written as
\begin{align}\label{pwh}
U(\rho,z;d) = -\dfrac{1}{16 \pi^{2}\varepsilon_{0}}(\Xi_{\rho}\braket{\mu_{\rho}^{2} } + \Xi_{\phi}\braket{\mu_{\phi}^{2} } + \Xi_{z}\braket{\mu_{z}^{2} }).
\end{align}
The $\braket{\mu^{2}_{i}}$ are the expectation values of the $i$-th cylindrical component of the dipole moment operator, and the coefficients $\Xi_{i}$ are:
\begin{widetext}
\begin{align}
\Xi_{\rho} &= \dfrac{d \rho^{2}}{R_{+}^{5}R_{-}^{5}}\left( P^{2} - d^{2}z^{2}\right) + \dfrac{d^{3}}{6 R_{+}^{3}R_{-}^{3}} + \dfrac{1}{4 z^{3}}\left[ \dfrac{\pi}{2} + \arctan\left(\dfrac{P}{d z}\right) + \dfrac{d z}{R_{+}^{4}R_{-}^{4}}Q_{-}^{2} P \right],\\
\Xi_{\phi} &=  \dfrac{d^{3}}{6 R_{+}^{3}R_{-}^{3}} + \dfrac{1}{4 z^{3}}\left[ \dfrac{\pi}{2} + \arctan\left(\dfrac{P}{d z}\right) + \dfrac{d z}{R_{+}^{2}R_{-}^{2}}P \right],\\
\Xi_{z} &= \dfrac{d}{R_{+}^{5}R_{-}^{5}}\left( z^{2}Q_{+}^{2} - \dfrac{d^{2}}{4}Q_{-}^{2}\right) + \dfrac{d^{3}}{6 R_{+}^{3}R_{-}^{3}} + \dfrac{1}{2 z^{3}}\left[ \dfrac{\pi}{2} + \arctan\left(\dfrac{P}{d z}\right) + \dfrac{d z}{R_{+}^{2}R_{-}^{2}}Q_{-} + \dfrac{2 d \rho^{2} z^{3}}{R_{+}^{4}R_{-}^{4}}P\right],
\end{align}
\end{widetext}
where we used the shorthand notations:
\begin{align}
P&= \rho^{2}+z^{2} -\dfrac{d^{2}}{4}\\
Q_{\pm}&=\rho^{2}\pm z^{2} \pm\dfrac{d^{2}}{4}\\
R_{\pm} &= \left[ \left(\rho \pm \dfrac{d}{2}\right)^{2} + z^{2}\right]^{1/2}.
\end{align}
For the hole diameter $d$ approaching zero, $U(\rho,z; d)$ reduces to a form $\propto$ $z^{-3}$--- a potential varying only in one direction, thus reducing to the familiar 1D form.  Equation~(\ref{pwh}) describes the energy shift of an atom with an arbitrarily oriented dipole; in our case, we choose the dipole to always be pointing in the direction of the atom's motion. We therefore parametrise the dipole moment as
\begin{align}
\boldsymbol{\mu_{\rho}} &= (\mu_{x} \cos \phi + \mu_{y} \sin \phi)\hat{\rho}  \\
\boldsymbol{\mu_{\phi}} &= (-\mu_{x} \sin \phi +\mu_{y}\cos\phi)\hat{\phi}\\
\boldsymbol{\mu_{z}}&=\mu_{z}\hat{z},
\end{align}
where $\hat{\rho},\hat{\phi},\hat{z}$ are the usual cylindrical unit vectors and $\boldsymbol{\mu} = (\mu_{x},\mu_{y},\mu_{z})$ is the dipole moment vector in Cartesian coordinates. By choosing a plane of motion where $\mu_{y} = 0$ and $\mu_{x} > 0$, we notice that $\phi = \arctan(y/x) = 0$. Now, by defining an angle $\theta = \arctan(x/z)$, we can write the remaining dipole components as $\mu_{x}=|\boldsymbol{\mu}|\sin\theta$ and $\mu_{z}=|\boldsymbol{\mu}|\cos\theta$. This allows us to write the energy shift $V(\rho,z;\theta,d)$ as
\begin{align}\label{pwh_or}
V(\rho,z; \theta, d) &= -\dfrac{1}{16 \pi^{2}\varepsilon_{0}}(\Xi_{\rho}\braket{|\boldsymbol{\mu}|^{2} }\sin^{2}\theta+ \Xi_{z}\braket{|\boldsymbol{\mu}|^{2}}\cos^{2}\theta)\nonumber \\
& = - \dfrac{C_{3}}{\pi}(\Xi_{\rho}\sin^{2}\theta + \Xi_{z}\cos^{2}\theta),
\end{align}
where $C_{3} = \braket{|\boldsymbol{\mu}|^{2}}/16 \pi \varepsilon_{0}$, with $\braket{|\boldsymbol{\mu}|^{2}}$ the square of expectation value of the dipole moment, and following \cite{C3Vals,Galiffi_2017_2D_sim}, we set $C_{3} = 4.0 \times 10^{-50}$J, which describes the interaction between ${}^{3}\mathrm{He}$ and a $\mathrm{Au}$ plate. By fixing the plane of propagation, $\rho$ and $z$  effectively become Cartesian coordinates, but for the sake of clarity and continuity we retain the cylindrical labels. The fixed, relative relationship between $\Xi_{z}$ and $\Xi_{\rho}$ components, such that the atom's dipole always points in the direction of motion is described by the angle of incidence $\theta$, and is schematically shown in Fig.~\ref{fig:pwhgraphic}.

\subsection{Extended potential}
\begin{figure*}
	\centering
	\includegraphics[width=0.7\linewidth]{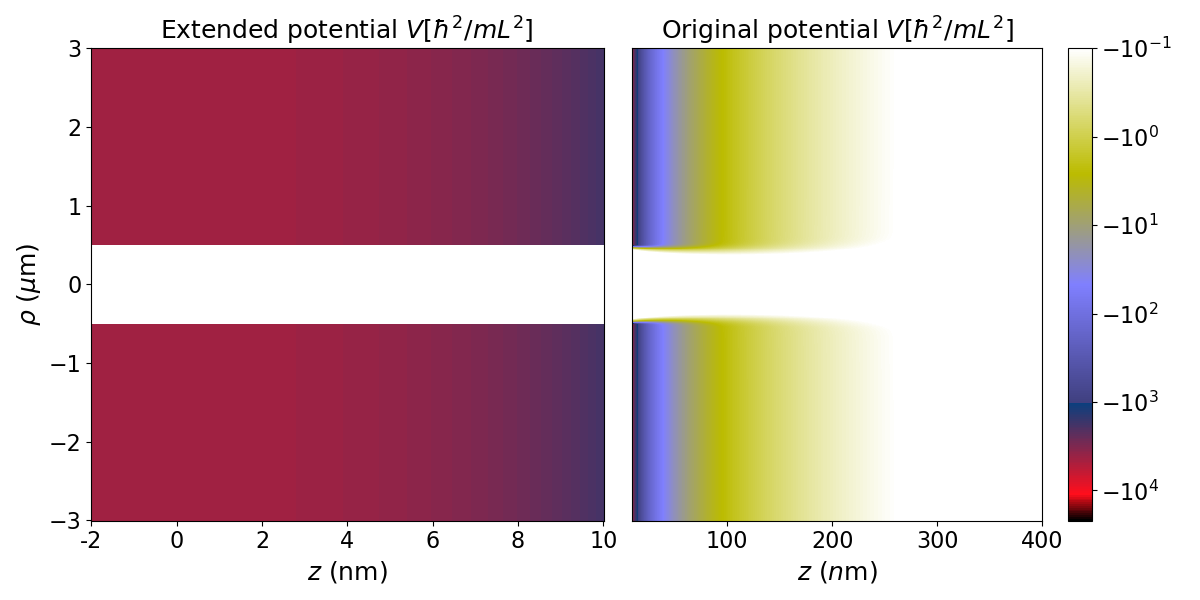}
	\caption{On the left, potential $V$ for an atom of $^{3}\mathrm{He}$, continued beyond the $z = \epsilon$ distance through to the negative $z$ values where it reaches a constant value. On the right, the behaviour of the original potential $V$ from the cut-off point at $z = \epsilon$. The unit $\hbar^{2}/m L^{2}= 0.014 \text{ neV}$.}
	\label{fig:potentialforms}
\end{figure*}

\begin{figure}
	\centering
	\includegraphics[width=1\linewidth]{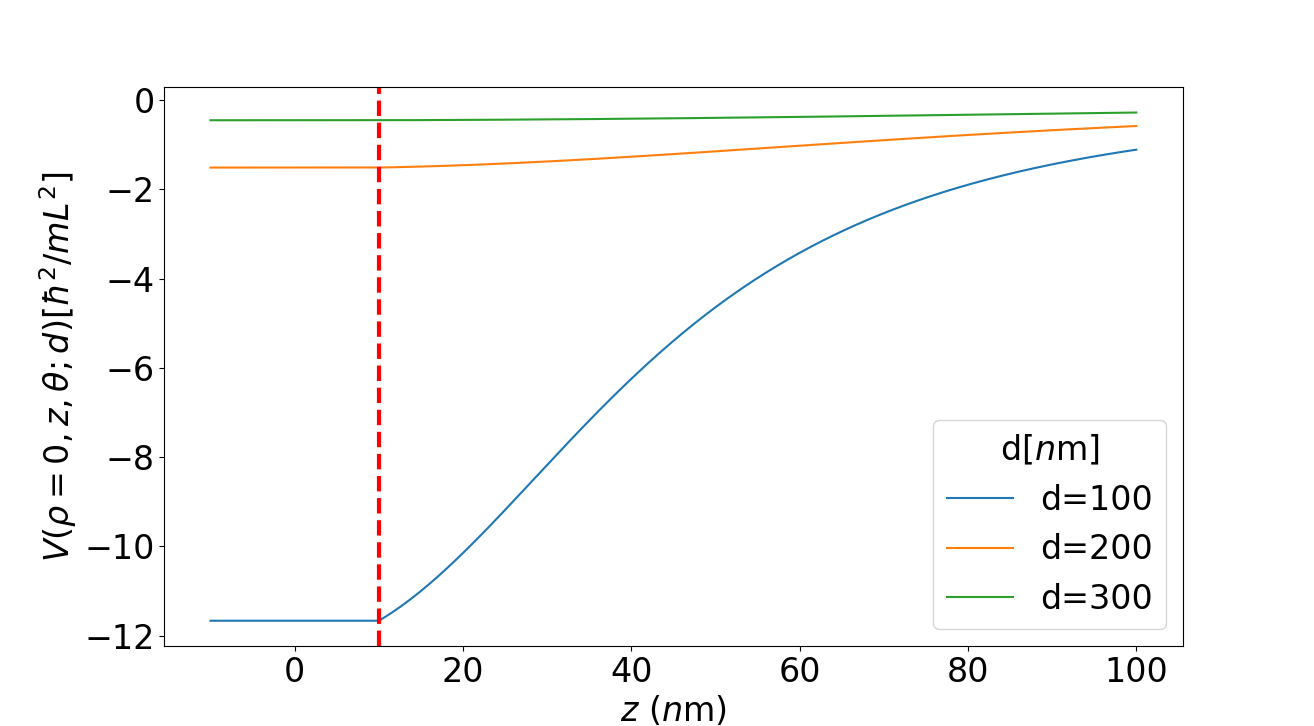}
	\caption{The extended potential $V_{C}$ at $\rho=0$ for three different hole diameters $d$. The red dashed line shows the cut-off point $\epsilon$. The larger the hole diameter, the flatter the potential gradient becomes near the centre.}
	\label{fig:potential1d}
\end{figure}

Any experiment aiming to measure quantum reflection needs to come up with a way of isolating it from the classical reflections induced by the short-range repulsion very close to the surface (for example using the bond dissociation technique in \cite{Dimer}). In our numerics we do this by simply not including the short-range repulsion (which would be implemented by letting $V\to\infty$ at the plate), and considering only the effects of the potential $V(\rho,z;\theta,d)$ as shown. This means that any reflections that occur are necessarily quantum in nature. To implement this we split our computational domain into two halves, with the plate envisaged as being in the middle at $z=0$. On the right hand side, the potential is $V(\rho,z;\theta,d)$, while on the left hand side we artificially continue the potential to the edge of the domain in the way we shall explain shortly.  Since the potential $V(\rho,z;\theta,d)$ experiences an unphysical singularity at $z=0$, we choose a small enough distance $\epsilon$, as a cut-off point (as was implemented, for example, in \cite{Oscillating_Herwerth_2013}). This length needs to be sufficiently small so the resulting potential still reaches close enough to the surface to be relevant to electrostatic interactions --- varying the $\epsilon$ impacts the reflectivity and this is discussed in the appendix where we test different lengths $\epsilon$. We now proceed to define a new piecewise potential function $V_{C}$ as 
\begin{align}
V_{C} = 
\begin{cases}
V(\rho,z,\theta;d)&  z > \epsilon \\\\
-\dfrac{3 V_{0}}{2 \epsilon^{2}}z^{2}+\dfrac{5 V_{0}}{2} &0 \le z \le\epsilon\text{  and  }  |\rho| <  \dfrac{d}{2}  \\\\
\dfrac{5 V_{0}}{2} &  z < 0 \text{  and  }   |\rho| < \dfrac{d}{2} \\\\
\quad V_{<} & \text{ otherwise },
\end{cases}
\end{align}
where $V_{0} \equiv V(0,\epsilon;\theta,0)$ and $V_{<}\equiv V(0,\epsilon;\theta,d)$.
The extended potential in the region $0\le z \le \epsilon$ is essentially a function of $z$ only. The change of the potential's landscape in the $\rho$ direction induced by introduction of the hole is symmetrical, and significant only at $z$ near $\epsilon$. We thus create a gap in the continued part of the potential in the positive and negative $\rho$ direction, at $z=\epsilon$, to account for the vanishing potential gradient at the hole's centre. For $z < \epsilon$ and beyond the gap, $V_{C}$ is invariant in $\rho$; for a wave packet travelling in the $\rho$ direction, such an abrupt change in the $\rho$ direction will have an effect on its motion. However, this is inconsequential for our purposes as this occurs in the continued part of of the potential and does not influence the reflection in the normal direction. We plot the regularized potential at a $\rho =0$ slice for different diameters $d$ in Fig.~\ref{fig:potential1d}.

\subsection{Evolution of the system}
We aim to solve a dimensionless time-dependent Schrödinger equation
\begin{align}\label{Schrod}
-\dfrac{1}{2}\nabla^{2}\Psi(\mathbf{r}, t) + V(\mathbf{r})\Psi(\mathbf{r}, t)=i\partial_{t}\Psi(\mathbf{r}, t).
\end{align}
We solve Eq.~(\ref{Schrod}) by taking advantage of an open source library \cite{Solver_Figueiras_2018} --- a solver utilising the split-step Fourier technique, also known as the Beam Propagation Method (BPM) \cite{BPMagrawal2007nonlinear}. Determined by the natural units and the choice of length scale $L = 1 \mu \text{m}$, the energy unit in which the system in Eq.~(\ref{Schrod}) is solved is $\hbar^{2}/m L^{2}$, where $m$ is the actual mass of the atom in SI units.  We adapt the source code of \cite{Solver_Figueiras_2018} to include our extended potential function $V_{C}$, and solve the TDSE for a range of chosen angles of incidence $\theta$, and diameters $d$. In Table \ref{tabb}, we specify the simulation specific parameters for $^{3}\mathrm{He}$ and Na, to which we refer throughout the text. At $t=0$, we define $\Psi(\mathbf{r}_{0},0)$ to be a Gaussian with $\sigma_{z}=\sigma_{\rho}= 1\mu\text{m}$, situated at the location $\mathbf{r}_{0}=\{ r\cos\theta, r\sin\theta\}$, where $\theta$ is the angle of incidence and  $r$ was chosen to be  $4\mu\text{m}$.  We impart on the pulse an initial momentum $p_{0} = \sqrt{2 m E_{0}}$, where $m$ is the mass of $^{3}\text{He} = 3.016$ $ \mathrm{ amu }$ ($\mathrm{Na}=22.99$ $ \mathrm{ amu}$), and $E_{0}$ is the kinetic energy. The computational domain is surrounded by an absorbing boundary, where the solver makes $V$ imaginary. Additionally, periodic boundary conditions are enforced and any pulse that ``leaks" through the absorbing medium, reappears at the starting point. We can stop that from happening if we choose a ``sensible" stopping point, an appropriate duration of propagation turns out to be $ t_{f}= 0.21$. The resulting $\Psi(\mathbf{r},t_{f})$ contains the information about the spread of the pulse at time $t_{f}$. To extract the information about the reflected part of the pulse, we simply integrate the normalised squared amplitude of the wave function along the $\rho$ axis, and the positive $z$. This way, we find the proportion of the pulse travelling in the positive $z$ direction at $t_{f}$, which we call reflectivity $R(t_{f})$,
 
\begin{align}\label{Refl}
R(t_{f})=\int_{-\infty}^{\infty} \mathrm{d}\rho\int_{0}^{\infty}\mathrm{d}z~|\Psi(\mathbf{r},t_{f})|^{2}.
\end{align}

An alternative, very similar treatment is to Fourier transform the $\Psi(\mathbf{r},t_{f})$ and integrate along the momentum in $\rho$ direction and positive momentum in $z$; the reader can follow \cite{Galiffi_2017_2D_sim,Oscillating_Herwerth_2013} for further details. We found this technique to produce almost identical results with the exception of cases where a pulse travels at grazing angles of incidence, causing the positive momentum to be poorly defined.

We notice that the addition of the hole significantly flattens the gradient of the potential in the region corresponding to its diameter across the whole domain, as seen in the right panel of Fig. \ref{fig:potentialforms}. This serves as basis for expecting suppressed reflectivity across those regions.

\begin{table}
	\centering
	\begin{tabular}{|c| c| c|}
		\hline
		 \rule{0pt}{2.5ex} % Add extra space below the top \hline
		Atom & $^{3}\text{He}$  & Na \\
		\hline
		Array dims.& $25\times25(25\times25$ $\mu\mathrm{m}$) &  $25\times25(25\times25$ $\mu\mathrm{m}$) \\ 
		Energy $E_{0}$ & $1.13\times10^{5}( 1.56 ~\mu\mathrm{eV}) $& $665.70( 1.21~\mathrm{neV})$ \\
		Time $t_{f}$& $0.21(2.21$ $\mathrm{ns})$ & $0.21(0.115$ $\mu\mathrm{s})$ \\
		Cut-off $\epsilon$ & $0.001(1$ $\mathrm{nm})$ & $0.1(100$ $\mathrm{nm})$ \\
		
		\hline
		
	\end{tabular}
	\caption{Parameters in natural units (SI units) used in the simulations.}
	\label{tabb}
\end{table}

\section{Results and analysis}\label{results}
\begin{figure*}
 	\centering
 	\includegraphics[width=1\linewidth]{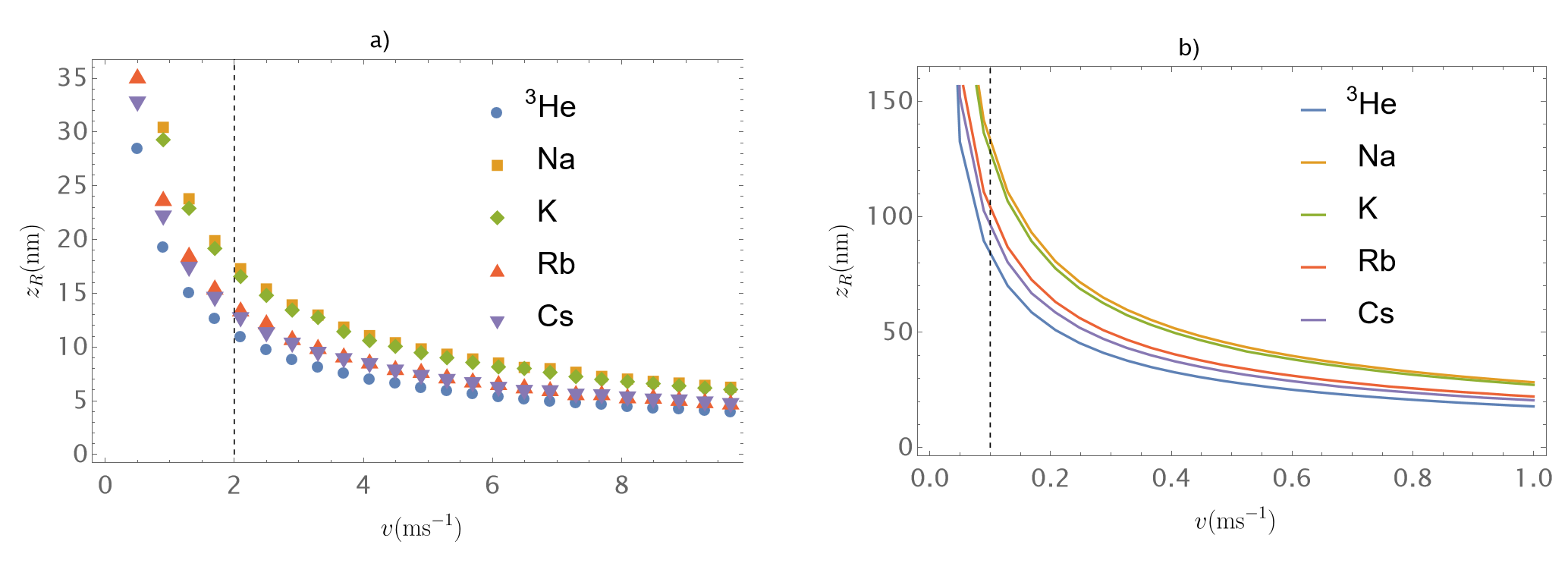}
 	\caption{The distance $z_{R}$ at which a quantum reflection occurs as a function of incident velocity, for various atoms. $z_{R}$ has been calculated as a location of the maximum of a Badlands function. The plot in a) shows the regime of applicability of the $^{3}\text{He}$ atom, with the dashed line marking its velocity $(\text{v}= 2 $ $\si{m.s^{-1}} )$; the same in b) but for the Na atom $(\text{v}= 0.1$ $\si{m.s^{-1}} )$.}
 	\label{fig:atoms}
 \end{figure*}

 Since we have based our investigations on using an electrostatic potential, we need to confirm that the reflection is happening at distances appropriate to the short-range, non-retarded CP interaction. The electrostatic regime is usually accessible through high kinetic energies where the particle's speed $v \approx 300$ \si{m.s^{-1}} --- as shown experimentally in for example \cite{exp2}. We can also consider non-retarded distances in the case of lower energies ($v \approx 2$ \si{m.s^{-1}}), by balancing out other parameters, i.e., choosing the $C_{3}$ coefficient (corresponding to a different atom for a case of a perfect reflector, or a combination of an atom and a surface for a more general treatment) to be sufficiently smaller than in the high-energy case. We can confirm the suitability of our setup to an electrostatic regime by reducing our problem to a single dimension --- normal incidence at $\rho=d=0$ --- and examining the location of a quantum reflection. A well-known estimation of the order of magnitude of this distance (in one dimension) can be inferred from the so called Badlands function \cite{Berry_1972,WKB2}, as demonstrated for example in \cite{gab1} and \cite{gab2}. The dimensionless form of the Badlands function $Q(z)$ can be written as
 \begin{align}
 	Q(z) = \frac{4(V(z)-E)V''(z) - 5(V'(z))^{2}}{32(E-V(z))^{3}},
 \end{align}
  where $V(z) \equiv V(0,z,0;0)$ is the one dimensional potential function, $E$ is the kinetic energy, and primes denote differentiation with respect to $z$.
  The peaks of $Q(z)$ coincide with regions where the WKB approximation breaks down (distances at which the wave vector experiences drastic changes), revealing the approximate position at which the quantum reflection occurs. Thus, by finding the location of a maximum of the Badlands function for a given configuration (choice of an atom and its velocity), we can check the applicability of a given regime. We have found the peaks of the Badlands function for the case of a perfect reflector for $^{3}\mathrm{He}$ along with $\mathrm{Na},\mathrm{K},\mathrm{Rb}$ and $\mathrm{Cs}$, using the $C_{3}$ coefficients for the alkali metals from \cite{atomsPhysRevLett.82.3589}. The results are shown in Fig. \ref{fig:atoms} a) and b). For all elements in Fig.~\ref{fig:atoms}  a), we notice a rapid growth of the distance $z_{R}$, for $v < 2 $ \si{m.s^{-1}}, a retarded regime corresponding to the usually associated with quantum reflection, lower energies. On the opposite side of the spectrum ($z_{R} \gg \lambda$, for a dominating transition wavelength of $^{3}\mathrm{He}$,  $\lambda = 9.3$ nm \cite{Oscillating_Herwerth_2013}), the distance $z_{R}$ falls inside the electrostatic (non-retarded) limit. We are thus considering a reflection distance which is approximately on the order of a wavelength $\lambda$, motivating us to discard any contribution that might be arising over the scale of retarded distances $(z\gg \lambda)$. We believe this to be a marginally justifiable assumption for $^{3}\rm{He}$, following the work in \cite{Galiffi_2017_2D_sim}, whose use of the electrostatic potential influences our own method. Additionally, the lowest limit for the cut-off point $\epsilon$ producing convergent results for all angles $\theta$ was found to be $\epsilon = 10 $ nm. This situates it within the approximate region where the Badlands function predicts the reflection to occur, yet it still allows for the full interaction to play out - the pulse starts reversing its motion before the cut-off point. For the case of the alkali atoms (K,Rb,Cs)  along with Na as shown in Fig.~\ref{fig:atoms} b), the distance at which the quantum reflection occurs, falls clearly in the non-retarded regime. In the case of Na, the distance $z_{R}$ is approximately five times smaller than its transition wavelength, $\lambda \approx 590 $nm.  
  
\begin{figure}
 	\centering
 	\includegraphics[width=1\linewidth]{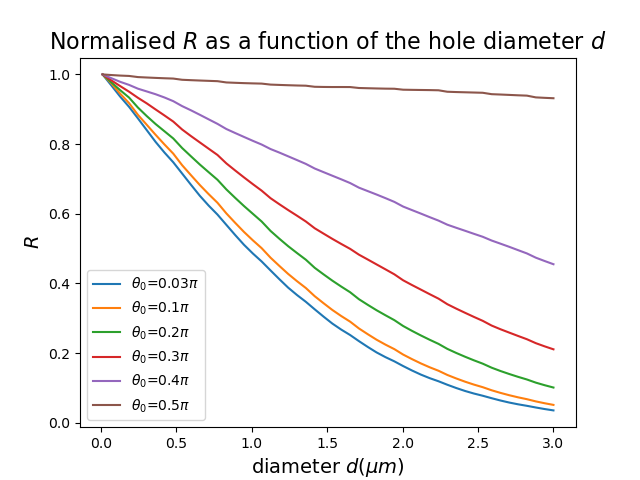}
 	\caption{Explicit dependence of reflectivity on the diameter of the hole $d$ for a selection of angles $\theta$ for the atom of $^{3}\mathrm{He}$ travelling at $\text{v}= 2$ $\si{m.s^{-1}} )$.}
 	\label{fig:refpoints50}
 \end{figure}

\begin{figure}
 	\centering
 	\includegraphics[width=1\linewidth]{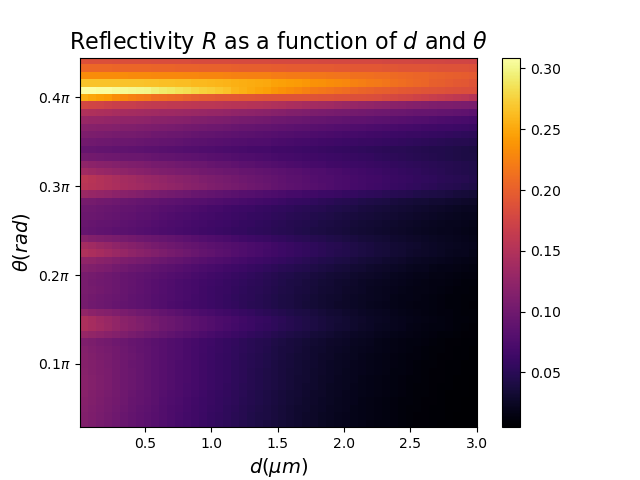}
 	\caption{Reflectivity as a function of the diameter of the hole $d$ and the angle of incidence $\theta$ for an atom of $^{3}\mathrm{He}$ travelling at $v= 2$ $\si{m.s^{-1}} $.}
 	\label{fig:ref2d50}
 \end{figure}

\begin{figure}
	\centering
	\includegraphics[width=1\linewidth]{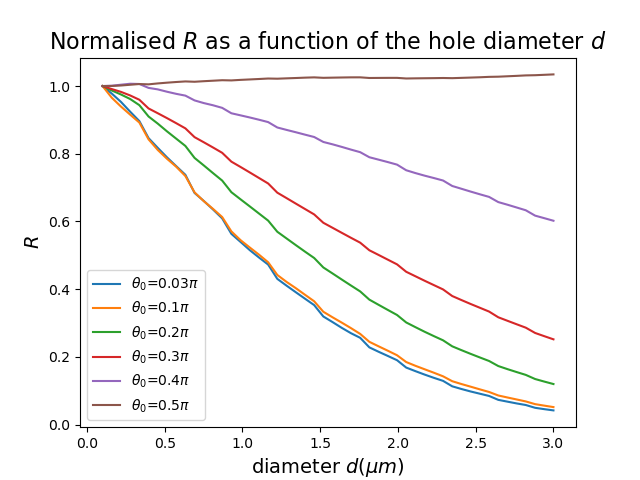}
	\caption{Explicit dependence of reflectivity on the diameter $d$ for an atom of Na travelling at $v = 0.1$  $ \si{m.s^{-1}}$.   }
	\label{fig:refpoints50naall}
\end{figure}

\begin{figure}
	\centering
	\includegraphics[width=1\linewidth]{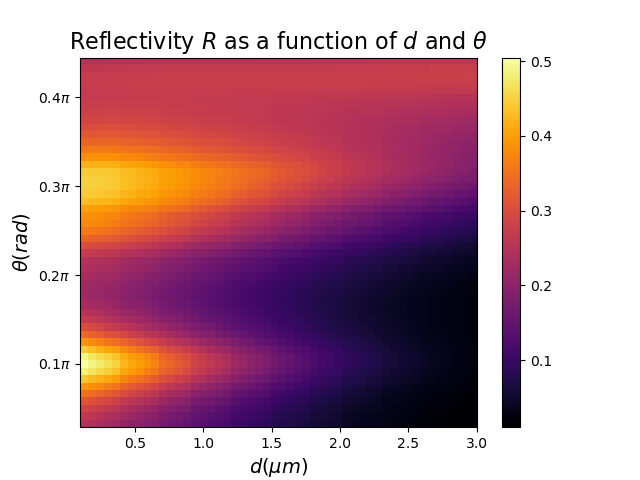}
	\caption{Reflectivity as a function of the diameter of the hole $d$ and the angle of incidence $\theta$ for an atom of Na travelling at $v = 0.1$ $\si{m.s^{-1}}$.}
	\label{fig:ref2d50naall}
\end{figure}

Having confirmed the validity of the one-dimensional electrostatic model for several atoms, we carry out two-dimensional simulations for an atom of $^{3}\text{He}$, and Na; in both cases we vary the incidence angle $\theta$ and the hole diameter $d$. The normalised (with respect to the initial reflectivity) results for $^{3}\text{He}$ (Na) are plotted in Fig.~\ref{fig:refpoints50} (Fig.~\ref{fig:refpoints50naall}), showcasing the relationship between the hole diameter $d$ and reflectivity $R$. Quite intuitively, for an atom travelling at  incidence angle $\theta$, the bigger the overlap between the hole's cross section and the arc that the angle $\theta$ subtends, the larger portion of the pulse experiences the reduced strength of the potential gradient. This can be clearly seen by the diminishing influence of the hole on pulses that travel at the grazing angle of incidence---the results of such simulations are shown in Figs \ref{fig:ref2d50}  and \ref{fig:ref2d50naall}. 

  Additionally, independent of the hole diameter $d$, in the case of each atom we observe a periodic behaviour along the $\theta$ axis.  The ratio of the reflected wave to the incoming one is modulated by the coupling between the potential's respective dependencies on $\rho$ and $z$. Curiously, when the diameter of the hole approaches zero ---which nullifies non-perpendicular dependence--- we still observe the periodic behaviour. Since this occurs for both atoms, we have examined the animations of the respective simulations and have found them to be describing the correct values of the reflectivity ---we expand on this point in the appendix. We suggest that the reason behind this phenomenon lies in the self-interference of the wave packet. As it strikes the potential barrier, it disperses in all directions, ultimately affecting the reflectivity in a quasi-periodic fashion. 
 
\section{Conclusions/summary}
We have presented a proof-of-principle method of controlling the magnitude of quantum reflection of a $^{3}\mathrm{He}$ atom from a perfectly reflecting plate by adding a circular hole of varying diameter at its centre. The addition of the hole significantly modifies the potential experienced by the atom and directly influences the probability of quantum reflection. We extended the parameter space familiar from standard quantum reflection approaches by allowing our matter-wave to travel at arbitrary incidence angles with respect to the surface. This introduces complications as the lack of a chosen single trajectory impacts the choice of boundary conditions in the time independent approach, rendering defining a suitable-for-all simulation space computationally infeasible. We thus have modelled the problem as a 2D pulse propagation in the presence of an attractive potential, and solved a TDSE using a split-step method, utilising an open source solver \cite{Solver_Figueiras_2018}. 

 Our results confirm the intuitions insofar as the increase in the hole diameter reduces the probability of the reflection---this is additionally influenced by the coupling between the direction of propagation and the strength of the potential gradient. The ability to study the reflection from the perspective of different directions of propagation reveals varied and interesting behaviours for the same atom. In the appendix we show how the finer grid density leads to convergence for the case of a normal incidence, and this will naturally apply to arbitrary direction of propagation. 
 
 The length scale at which we tested quantum reflection is ideally suited to the regime of nanotechnology, opening up possibilities for designing tunable quantum reflection devices, such as velocity selectors able to filter out neutral atoms \cite{vel_sel}.  As well as the range of possibilities in technological applications, the plate with the hole offers an interesting scheme for investigating quantum nature of matter waves. In this paper, we have discussed the behaviour of a single atom incident on the perfectly reflecting surface, but the same method (perhaps at lower energies and thus considering retarded distances) can be applied to studying the quantum reflection of a BEC, with the specific emphasis on the two-dimensional profiles, which will be explored in a future work. Alternative avenues exist to extend and interpolate the plate with the hole potential to a non-electrostatic regime in the form of a heuristic argument as it is often done in dispersion force calculations \cite{Shimizu2001}, or numerical simulations. Both remain to be respectively tested to expand the reach of possible quantum reflection experiments.       
\begin{acknowledgments}
It is a pleasure to thank Marc Caffrey for discussions. Financial support from UK Research and Innovation grant EPSRC/DTP 2020/21/EP/T517896/1 is gratefully acknowledged.
\end{acknowledgments}
\appendix*
\section{Convergence}\label{convergence}
As already pointed out by Galiffi et al. \cite{Galiffi_2017_2D_sim}, the convergence of a solution to the 2D pulse propagation problem depends on the density of points along the axis of the particle's propagation. They report using different grids for $x$ and $y$ values ---  the pulse is travelling only along the $x$ axis (normal incidence). In our case of arbitrary incidence, shortening the grid in the $y$ direction leads to spurious results, i.e., the direction of the pulse acquiring a phase of $- 2\theta$, where $\theta$ is the angle of incidence. As there is no preferred direction of motion, we thus use grids that have equal density across $z$ and $\rho$. We have inspected the animations of our simulations to establish a lower bound on the number of grid points $N$, for which the pulse follows a correct trajectory and we have found it to be $2^{11}$. Furthermore, we tested more dense grid configurations of the form $n\times n$ and were limited by memory to the case of $N=2^{13}$. Thus, we performed the numerical simulations --- results of which are shown in Figs: \ref{fig:refpoints50}, \ref{fig:ref2d50}, \ref{fig:refpoints50naall} and \ref{fig:ref2d50naall} --- using the $z$ grid of $N_{z} = 2^{12}$, balancing accuracy and performance. Moreover, the algorithm of the split step numerical method converges for small values of t \cite{Solver_Figueiras_2018} --- in our case the time step is chosen to be $dt = 0.005$.

\begin{figure*}
	\centering
	\includegraphics[width=1\linewidth]{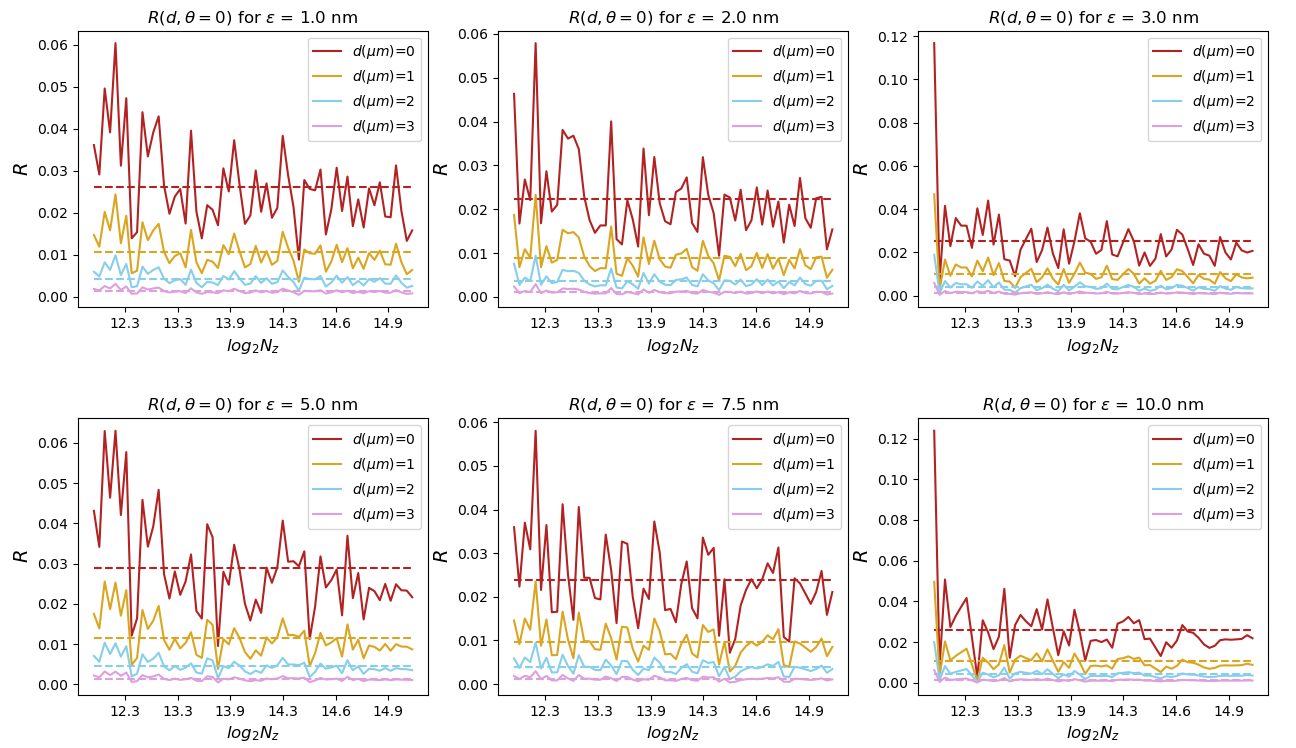}
	\caption{Relationship between reflectivity $R$ and grid density in the $z$ direction for normal incidence for the $^{3}\mathrm{He}$ atom. The number of points on the $\rho$ axis is kept constant, $N_{\rho}=2^{7}$. Different coloured lines correspond to different diameters of the hole --- which when increased reduce reflectivity as well as the magnitude of the fluctuations.}
	\label{fig:sixplots}
\end{figure*}

It is worth noting that introducing the regularization of the potential in the form of a cut-off length $\epsilon$ has an influence on overall results.  With decreasing $\epsilon$, the potential gradient a particle is experiencing becomes larger, and a denser grid is needed for more accurate sampling. We have tested this relationship using our algorithm for the case of normal incidence for different hole diameters, as shown in Fig. \ref{fig:sixplots}. The number of points on the $\rho$ axis was fixed to $N_{\rho}=2^{7}$, and we varied the density in the $z$ direction between $N_{z} = 2^{11}$ and $N_{z} = 2^{15}$. The values of cut-off length $\epsilon$ are bound by the reflection distance $z_{R}$, and were chosen between $1$ and $5$ nm --- shown as separate panels in Fig. \ref{fig:sixplots}. In each case, we observe that the amplitude of fluctuations around a mean value (dashed line) decreases as the number of points is increased. For increasing diameter $d$, the oscillations also decrease; the presence of the hole weakens the magnitude of the gradient in the normal direction. Thus, even a smaller resolution is able to capture the behaviour adequately. For our choice of range of $\epsilon$, the oscillations decrease in a similar manner until $\epsilon = 10$ nm, where they become more smoothed out for $N_{z}>2^{14}$.

All numerical computations were performed on a PC with an 8 core 11th Gen Intel(R) Core(TM) i7-11700 @ 2.50GHz, 16GB of RAM, and a Rocky Linux operating system.

\section{Periodic behaviour}
\begin{figure*}
	\subfloat[]{%
		\includegraphics[width=.3\linewidth]{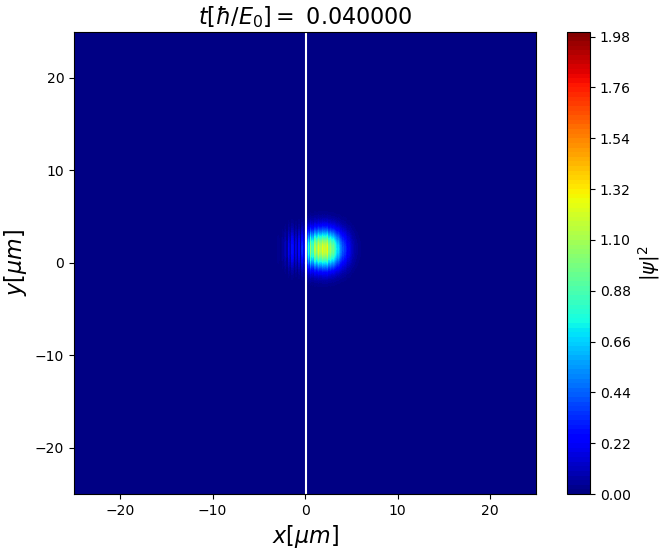}%
	}\hfill
	\subfloat[]{%
		\includegraphics[width=.3\linewidth]{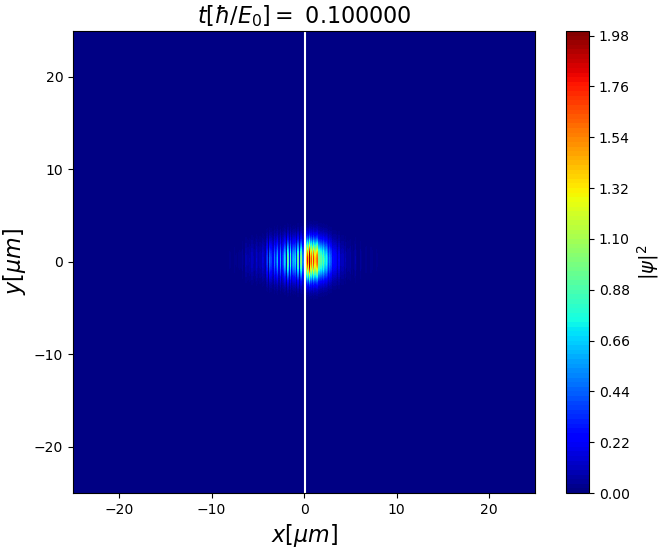}%
	}\hfill
	\subfloat[]{%
	\includegraphics[width=.3\linewidth]{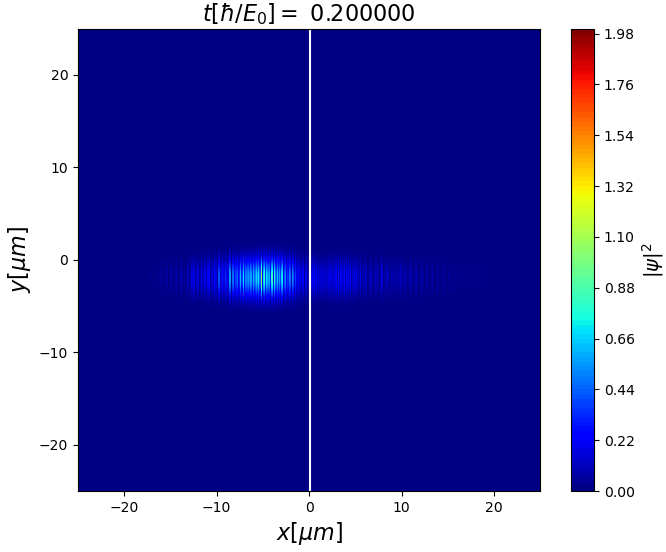}%
	}
	
	\subfloat[]{%
		\includegraphics[width=.3\linewidth]{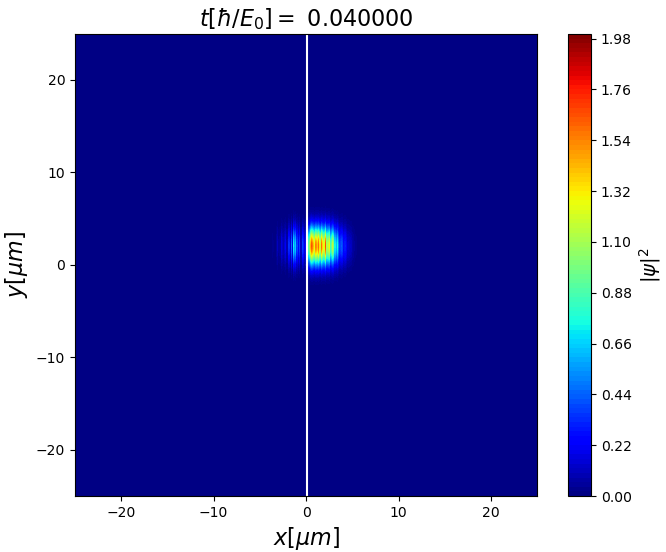}%
	}\hfill
	\subfloat[]{%
		\includegraphics[width=.3\linewidth]{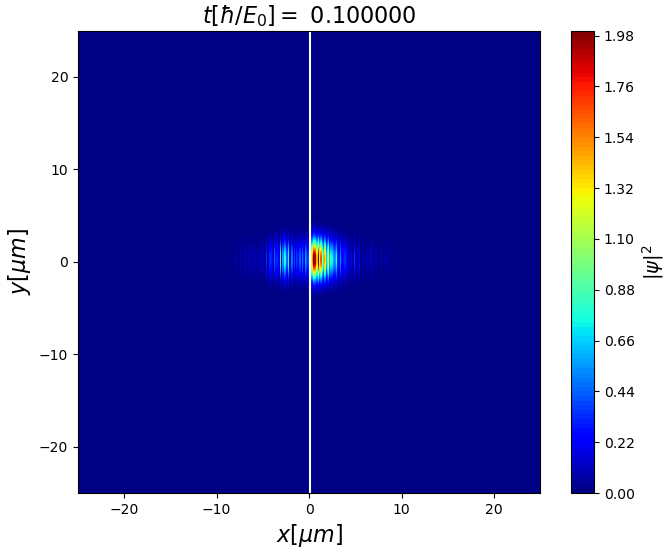}%
	}\hfill
	\subfloat[]{%
		\includegraphics[width=.3\linewidth]{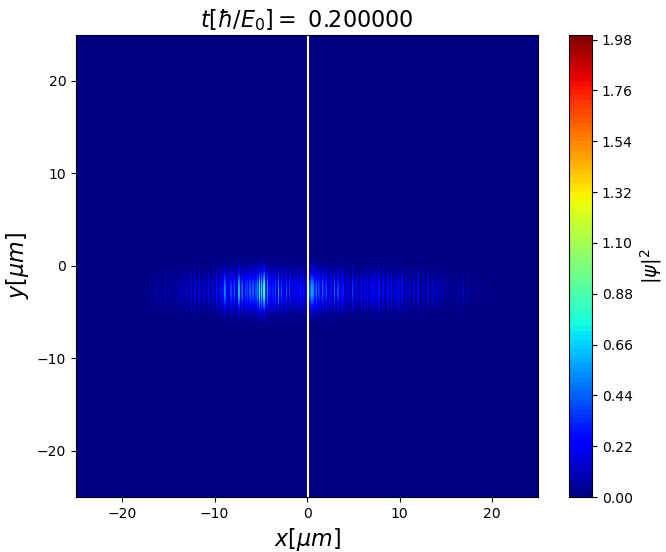}%
	}
	\caption{Different stages ($t = 0.04, 0.1, 0.2$) of the wave-packet propagation and scattering for the Na atom at a plate without a hole. The top row shows low reflectivity---resultant from propagation at angle $\theta=0.2\pi$, whereas the bottom row depicts high reflectivity--- angle $\theta=0.3\pi$. The standard deviation $\sigma_{\rho}$ and $\sigma_{z}$ have been increased to $2$ for ease of distinguishing the wave packet features.   }
	\label{vistest}
\end{figure*}

We have examined the animations produced by the simulations and found the visual representation to agree with the calculated values of reflectivity. The plots can be seen in Fig.~\ref{vistest}, there, we have included snapshots from the simulations of the Na atom where angles of incidence were respectively $\theta = 0.2\pi$ and $\theta=0.3\pi$. The respective resultant reflectivities $R(t_{f})$ were $0.221$ and $0.353$, which agree with the main results, leading us to assume that the reflectivity calculations are correct for all $\theta$. The influence of $\theta$ on periodic behaviour seen in Figs \ref{fig:ref2d50} and \ref{fig:ref2d50naall} cannot be explained through the action of simple functions such as $\sin\theta$ ($\cos\theta$) since they are strictly increasing (decreasing) on the interval $(0,\frac{\pi}{2})$. Thus, a more complicated response must be at play, being born out of the scattering of the wave packet across different angles of incidence. Given the strong non-separability of (\ref{Schrod}), we are unable to investigate this behaviour analytically. 
\clearpage  
\bibliography{QR_pub.bib}
\end{document}